\documentclass{article}
\usepackage{latexsym}
\usepackage{graphicx}
\usepackage{mathptmx}

\usepackage{amsmath}
\usepackage{amsfonts}
\usepackage{amssymb}
\usepackage{amsbsy}
\usepackage{amsthm}

\usepackage[pdftex,colorlinks=true,urlcolor=blue,citecolor=black,anchorcolor=black,linkcolor=black]{hyperref}

\usepackage{cite}
\usepackage{tikz}
\usepackage{tikz-qtree}
\usepackage{fullpage}
\usetikzlibrary{shapes,arrows,positioning, calc}

\tikzstyle{decision} = [diamond, draw, fill=yellow!20, 
    text width=4.5em, text badly centered, node distance=3cm, inner sep=0pt]
\tikzstyle{block} = [rectangle, draw, fill=blue!20, 
    text width=5em, text centered, rounded corners, minimum height=4em]
\tikzstyle{subroutine} = [draw,rectangle split, 
	 rectangle split horizontal,rectangle split parts=3,
	 fill=blue!20, minimum height=4em]
\tikzstyle{line} = [draw, -triangle 45]

\tikzstyle{data} = [draw,trapezium,trapezium left angle=80,text badly centered, text width=3em,trapezium right angle=-80,fill=green!20,minimum height=4em]

\tikzstyle{stop} = [draw, ellipse,fill=red!20, node distance=3cm,
    minimum height=4em]
        
\newtheoremstyle{wsc}
{3pt}
{3pt}
{}
{}
{\bf}
{}
{.5em}
{}

\theoremstyle{wsc}

\begin{document}

%
%

\title{A COMPOSITE CONSTRAINTS APPROACH TO DECLARATIVE AGENT-BASED MODELING}

\author{
David Bruce Borenstein \\
\vspace{12pt} \\
Lewis-Sigler Institute for Integrative Genomics \\
Princeton University \\
Carl Icahn Laboratory \\
Princeton, NJ 08544, USA \\
}

\maketitle

\section*{ABSTRACT}
Agent-based models (ABMs) are ubiquitous in research and industry. Currently, simulating ABMs involves at least some imperative (step-by-step) computer instructions. An alternative approach is declarative programming, in which a set of requirements is described at a high level of abstraction. Here we describe a fully declarative approach to the automated construction of simulations for ABMs. In this framework, logic for ABM simulations is encapsulated into predefined components. The user specifies a set of requirements describing the desired functionality. Additionally, each component has a set of consistency requirements. The framework iteratively seeks a simulation design that satisfies both user and system requirements.  This approach allows the user to omit most details from the simulation specification, simplifying simulation design.

\section{INTRODUCTION}
\label{sec:intro}

\subsection*{The need for descriptive modeling}

Agent-based models (ABMs) constitute one of the most widely used categories of simulation technology. ABMs (also known as ``individual-based models'' or IBMs) represent a system as an ensemble of autonomous actors (or ``agents''). These agents interact with one another according to predefined behaviors. The set of defined behaviors may be unique to each individual agent, or common to a class of agents. ABMs are widely used for academic research in the fields of ecology, epidemiology and social science \cite{Eubank2004,Grimm2005,Gilbert2008}. Commercial and governmental uses include business analytics, supply chain management, and civil and military planning \cite{Fox2000,Lucas2004,Delre2007,Zheng2009}.

ABMs provide a link between local and global dynamics. The modeler defines local interactions. As these interactions play out, global patterns often become evident. By choosing local rules based on observed real-world processes, ABMs enable the modeler to predict large-scale consequences. Conversely, by selecting rules that recapitulate observed large-scale processes, modelers can make predictions about the underlying local interactions. In either case, effective use of ABMs requires deep insight into the process at hand---a body of knowledge wholly disjoint from the computer expertise required to actually build models. These challenges are all exacerbated by the introduction of spatial structure, where topological details can have major implications for emergent behavior \cite{Durrett1994,Borenstein2013}.

The predominant representation of ABM outside of software is a standardized rubric of model features, called the `Overview, Design concepts, and Details'' (ODD) approach \cite{Grimm2006}. In an ODD specification, agent-based models are described in terms of their structure and temporal dynamics. Notably, computer logic is minimized in the ODD specification:  the rubric focuses on what the model does, rather than how a programmer chose to accomplish it. Simultaneously, there have been efforts to develop a logic-based descriptive framework for the general description of simulations \cite{Guizzardi2010}. Here, we present a declarative agent-based simulation framework, Nanoverse, that is based on the principle of description. 

In existing approaches, model design is closely linked to simulation design. The model deals with the properties of the agents and the world they occupy. When do they act? What information can they incorporate into their choices? The simulation, on the other hand, is a computer program capable of actualizing the model and integrating it over time \cite{Miller2007}. For the most part, existing tools are simulation frameworks: it is up to the user to first envision a model and then articulate a process for simulating it. One exception is the commercial tool Anylogic, discussed below, which uses flowcharts to focus attention on the model. In any case, the user must still define a sequence of computer instructions.

The key innovation behind Nanoverse is that it structures ABM implementation as a configuration problem. Rather than specify step-by-step rules, the user imposes constraints (requirements) on the model. The platform then uses this specification to find a configuration of predefined components that satisfies these constraints. By replacing step-by-step (imperative) computer instructions with a (declarative) description of a model's properties, simulation design can be brought closer in line with the ways in which ABMs are discussed. 

\subsection*{Imperative approaches to ABM design}

Simplifying ABM development has been the focus of much research and development. Much of this research has focused on general-purpose software tools for spatially structured ABMs. Most ABM software tools have introduced expressive computer languages (or language extensions) created for the specific purpose of ABM simulation \cite{Railsback2006}. By far the most successful project has been NetLogo, which extends the educational programming language LOGO \cite{Feurzeig1969} to a large library of agent-specific structures and actions. NetLogo \cite{Wilensky2004} has been widely adopted in academic research, and remains popular after nearly two decades of continuous use. An extreme form of this approach is purely visual programming, as in StarLogo TNG \cite{Klopfer2007} and Scratch \cite{Maloney2010}. These K-12 educational tools make modeling easier by representing imperative statements as visual blocks.
  
Paradoxically, the simplicity of these LOGO-derived tools means that complex models can be challenging to express. Another tool, GAMA \cite{Grignard2013} seeks to address some of these limitations by providing straightforward facilities for GIS and multi-level models. GAMA utilizes a fluent, object-oriented language called GAML. GAML automates many aspects of model design, but still requires the user to specify and manipulate data structures. 

Other imperative tools for complex simulations include Java libraries such as MASON \cite{Luke2005} and Repast \cite{North2013}. These tools each provide a powerful, object-oriented framework within which to build and simulate ABMs. However, these tools require proficiency with general-purpose programming languages such as Java. Commerical package AnyLogic uses a variety of UML-like charts to represent the states and actions of agents, which it then translates to Java code \cite{Borshchev2002}. While highly accessible, this approach is essentially analogous to imperative programming since it requires the user to define a sequence of logical actions.

\subsection*{A component-based architecture for ABMs}

Many agent-based models can be simulated using a common set of strategies. NetLogo, GAMA, Repast and StarLogo all provide an extensive library of common logical pieces; often the only programming task is to unite these pieces in a manner appropriate to the model. Nanoverse extends this concept further, by hierarchically building up components from a pool of subcomponents. By repeatedly applying this idea, it is possible to define agent-based models from a relatively small body of simple units. Since all imperative logic would be encapsulated in these units, the user's task becomes one of describing conceptual relationships, rather than computational tasks. This is the principle behind component-based (or ``modular'') software engineering \cite{Bachmann2000,Reinhold2014}.

There is precedent for a component-based approach to simulation: SimKit provides a structure for building and distributin reusable imperative blocks, which can then be composed programatically or visually \cite{Buss2002,Buss2002a,Buss2007}.To the author's knowledge, there has been no effort to leverage the declarative nature of component-based software in order to present simulation design as a configuration task. This approach opens up a wealth of existing strategies for simulation implementation, as configuration problems are a cornerstone of knowledge engineering \cite{Wielinga1997}.

Configuration problems can be solved using constraint satisfaction approaches. In a constraint satisfaction problem, a ``solution'' is any value which satisfies every specified constraint. In configuration problems, constraints are either directly specified by the user, or are a consequence of a partial solution. For example, suppose that the user specifies a spatially explicit, two-dimensional model with periodic boundary conditions. It follows that the arena must have four sides. The goal of a constraint-based configuration scheme is to satisfy both the requirements imposed by the user and the requirements imposed by the selected sub-components, given a set of available options.

A straightforward approach to constraint satisfaction is backtracking. In a backtracking algorithm, solutions are tested sequentially against the first constraint, being globally eliminated if they violate it. Once a solution is found, the algorithm recurs on the next constraint. If no solution satisfies a constraint, it ``backtracks'' to the previous constraint, which resumes its search \cite{Wielinga1997}. By specifying a sequence of default subcomponents, the backtracking strategy is sufficient to configure a single component of a simulation, such as a spatial structure.

An entire simulation can be specified by nesting constraint satisfaction problems together, in a strategy known as composite constraint satisfaction (CCS) \cite{Sabin1996}. For example, suppose an agent may move or replicate into a nearby location. Are collisions with other agents allowed? If so, how are they resolved? If not, what happens if no legal move is possible? CCS configures these elements sequentially, either as nested sub-problems or as co-constraints within the same problem. This approach has previously been used to automate other software configuration tasks, such as the deployment of complex software systems \cite{White2007}.

This paper describes part of an ongoing effort to create a constraint-driven, spatially explicit agent-based modeling framework. This framework, called Nanoverse, is being prototyped in stages. The first stage, a working mock-up of which is available online \cite{Borenstein2015}, is a component-based simulation environment that is functionally similar to GAMA or MASON. This paper concerns the second stage of the prototype, currently under development, consisting of a multi-stage compiler. The paper begins with a brief synopsis of the runtime environment into which the compiler instantiates simulations. The second part describes the architecture of the Nanoverse compilation pipeline.

\section{RUNTIME SCHEME}

The Nanoverse runtime consists of a network of loosely coupled components. The primary subsystems of the runtime are a collection of topologies called ``layers'' and a discrete event scheduler \cite{Fishman2001}

The layer encapsulates all topological information, and the schedule encapsulates all scheduling information. Mutation of the simulation state is accomplished through scheduling events with a relative waiting time, which is subsequently resolved by the schedule. Likewise, specific changes to the environment are specified relative to a particular agent. As such, agents remain completely agnostic to the global state of the simulation.
	
In order to accomplish this, events have callbacks that request specific changes to their locale. Agents have a symbol table associating specific names with event triggers, as well as a rule table mapping specific conditions to the triggering of certain events. When a simulation event runs, it notifies the layer, which notifies all affected agents to consult their rule table. Forward integration of the simulation is achieved by repeatedly polling the schedule for the next scheduled event. The system time is advanced to that of the next event, and the event is allowed to run. If running the event triggers immediate events or alters the progression of future events, the schedule is updated. The simulation ends when the event queue is depleted or another terminal condition is met. The loose coupling of simulation components allows for the use of a constraints-based compiler system, which in turn allows us to move away from imperative programming.

\section{COMPILE SCHEME}

\subsection{Overview}

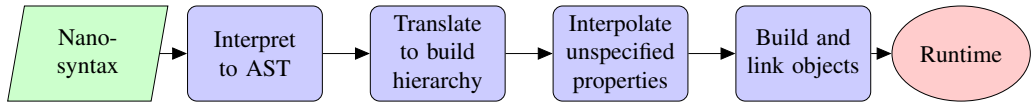
\begin{figure}[htb]
\centering
\scalebox{0.9}{
\begin{tikzpicture}[node distance = 2.7cm, auto]

        \node (nanosyntax) [data] at (0,0) {
            {Nano-
            \\syntax}
        };

        \node [block, right=0.4cm of nanosyntax] (interpret) {
            Interpret to AST
        };
        
        \node [block, right of=interpret] (translate) {
            Translate to build hierarchy
        };

        \node [block, right of=translate] (interpolate) {
            Interpolate unspecified properties
        };
        
        \node [block, right of=interpolate] (build) {
            Build and link objects
        };

        \node [stop, right=0.3cm of build] (runtime) {
            Runtime
        };

        \draw [line] (nanosyntax.east) -- (interpret.west);
        \draw [line] (interpret.east) -- (translate.west);
        \draw [line] (translate.east) -- (interpolate.west);
        \draw [line] (interpolate.east) -- (build.west);
        \draw [line] (build.east) -- (runtime.west);        

\end{tikzpicture}
}
\caption{Nanoverse compilation pipeline} \label{fig:pipeline}
\end{figure}

The Nanoverse compiler prototype consists of a four-stage compilation pipeline, ultimately leading to a discrete-event runtime for spatially explicit agent-based models (Fig. \ref{fig:pipeline}). The first stage of the pipeline is a parser that interprets a hierarchical source code into an abstract syntax tree (AST). This abstract syntax tree has no semantic information about the structure of an agent-based model; it reflects only the grammar of the user's specifications. The second stage uses a hierarchy of symbol tables to convert the abstract symbol tree into a semantically rich hierarchy of ``build nodes.'' These build nodes roughly correspond to the Java objects that will represent the simulation in memory. The third stage of the pipeline is the backtracking constraint solver, which is used to interpolate unspecified properties of the simulation. This is done by treating the user's specifications as additional constraints on an ordered sequence of defaults, with over- or underdefined specifications leading to an error. Finally, the completed build tree is visited breadth-first in order to instantiate all nodes into Java objects. The top-level object then triggers the execution of the simulation.

\subsection{The Nanosyntax environment}

The user writes Nanoverse model descriptions using a hierarchical grammar called ``Nanosyntax.'' Nanosyntax was influenced by the JSON object specification, which is used to serialize data for transmission between internet servers and clients \cite{EcmaInternational2013}. All statement blocks are terminated by a semicolon. Statements consist of an outer node, and, in the case of an assignment, one or more value nodes. Single value assignments are designated by a colon, and block assignments are specified by curly braces. The top level of a Nanoverse project specification is internally represented as a block assignment to a hidden reference. 

Nanosyntax consists of three types of nodes: ``primitives,'' ``references'' and ``assignments.'' Primitives are b​asic data types, such as numbers and strings. References specify an identifier or a block of identifiers. Assigmments map an identifier to a reference. Additional structures are allowed for mathematical operations, but these are converted internally to the other node types. These three elements are sufficient to specify an arbitrary hierarchy of members in a concise and readily intelligible way. 

Nanoverse project specifications consist of a nested ensemble of constraints, or explicit requirements concerning the properties of the simulation. Any requirements left unspecified are subsequently interpolated from a set of defaults to match the constraints that were specified.

The properties to be specified correspond directly to encapsulated operational units, or components, of the simulation's business logic. Thus, Nanosyntax is fully declarative: the only purpose of the source code is to describe what should be done, rather than how it should be implemented. As a result of interpolation, Nanosyntax is also minimal: the source code contains only the requirements of interest to the user. The user can therefore begin running simulations with very little code. By iteratively overriding defaults, the user can then build up the behavior of the simulation until it reflects all desired functionality.

As an example, consider the ``StupidModel'' reference model developed by \cite{Railsback2006}. The first of 16 instances of the model consists of a population of 100 agents that diffuse around a 100x100 rectangular lattice. Here we present the anticipated Nanosyntax for an even simpler model, which consists of a single agent diffusing around a 32x32 rectangular lattice. (Time is specified in arbitrary units.)

\vspace{4mm}
\begin{verbatim}
initially: 
    scatter:
        description:
            Agent:
                do: Behavior {
                    action: wander;
                    every: 1.0;
                    until: time >= 100.0;
                };
                
\end{verbatim}
\vspace{4mm}

\noindent The Nanosyntax representation of the model describes the entire system in a single statement block. Absent a specific geometry requirement, the system defaults to a 32x32 rectangular lattice with absorbing boundary conditions. An \verb|initially| assignment specifies one or more events that must occur to set up an initial condition.  Since the only action -- diffusion, or \verb|wander|ing---is encoded in the definition of the agent itself, there is no need to define a main loop. The \verb|wander| operation itself does have subcomponents dealing with destination selection and collision resolution, but these are also handled with interpolated defaults. At its construction time, the agent will schedule a \verb|wander| action that, upon firing, will schedule itself for one time unit later, until the time exceeds 100. By default, an image sequence of the process will be generated, with frames recorded at the start of each time unit. Specifying an alternative boundary condition, lattice geometry or arena shape would take one additional line apiece.

Existing frameworks require far more code to accomplish the same goal. MASON and Repast both require the user to define diffusion from first principles, instantiate a 2D arena, and place the agents using a random number generator; moreover, all of this must be done in Java. NetLogo eliminates the need for low-level programming, but still requires explicit instructions for each operation involved \cite{Railsback2006b}. GAMA requires that the user first define a geometry and a neighborhood structure, then the conditions for an ongoing behavior (or ``reflex'') representing movement. The user then defines a visual representation of the agent, and a display mode for the visual representation \cite{Amouroux2014}. The GAMA approach is similar to that of Nanoverse, except that Nanoverse is designed to resolve many of the specified details that are required in GAMA. 

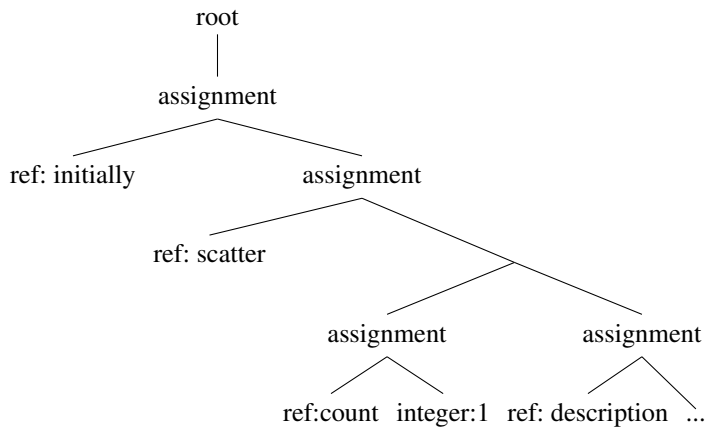
\begin{figure}[htb]
\centering
\begin{tikzpicture}[sibling distance=0pt]
\Tree [ .root [.assignment 
   [ .ref:\ initially ] [ .assignment [ .ref:\ scatter ] 
   [ [ .assignment [ .ref:count ] [ .integer:1 ] ] 
   [ .assignment [ .ref:\ description ] [ .... ] ] ] ] ] ]
\end{tikzpicture}
\caption{Top portion of an abstract syntax tree.} \label{fig:ast}
\end{figure}

The Nanoverse compiler parses Nanosyntax using the parser generator ANTLR4 \cite{Parr2014}. ANTLR4 generates a parse tree. The Nanoverse compiler then translates the Nanosyntax parse tree into an abstract syntax tree (AST), an example of which is shown in Fig. \ref{fig:ast}. Nanoverse employs a heterogeneous AST: different nodes are used for each of the three basic data types \cite{Parr2010}. By distinguishing between data types, the AST provides structural information that simplifies the next process in the pipeline: semantic analysis.

\subsection{Adding semantic information}

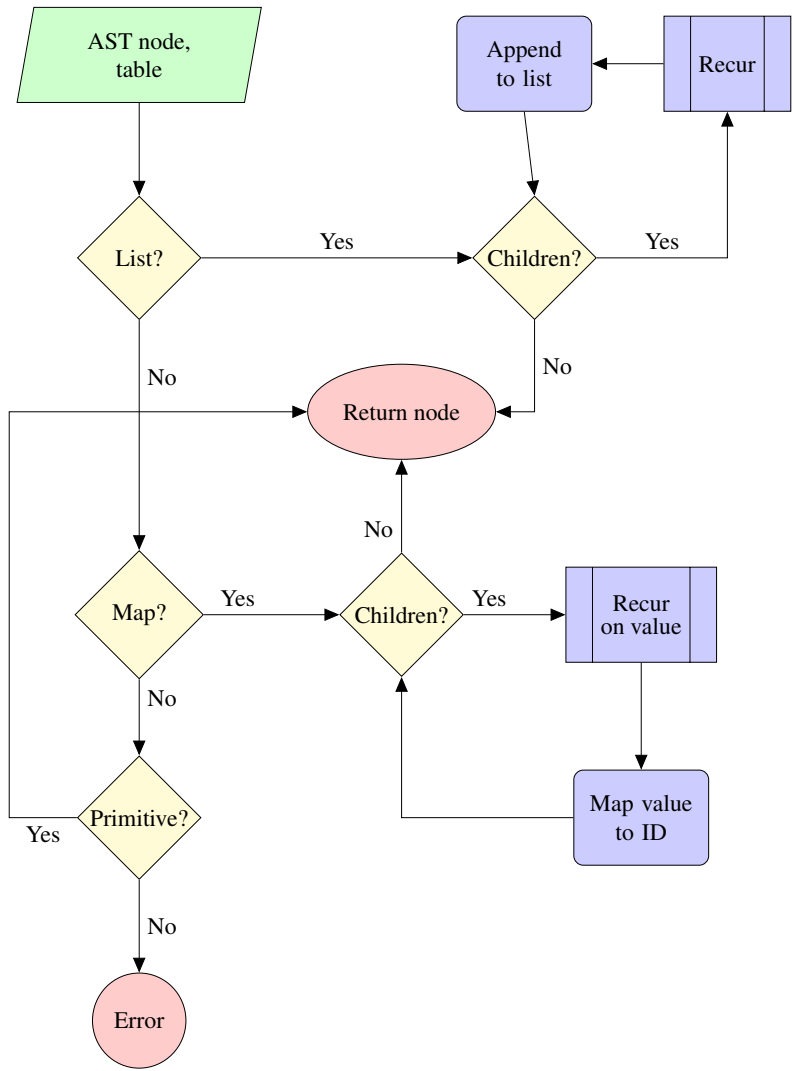
\begin{figure}[htb]
\centering
\scalebox{0.9}{
\begin{tikzpicture}[node distance = 3cm, auto]

		
        \node (input) [data, text width=5em] at (0,0) {
            AST node, table
        };

        \node [decision, below of=input] (list) {
            List?
        };

        \node [decision, right=4cm of list] (list-children) {
            Children?
        };


        \node [subroutine, above right=1.7cm and 1.445cm of list-children] (recur-list) {
            \nodepart{two}\shortstack{Recur}
        };

        \node [block, left of=recur-list] (append-list) {
            Append to list
        };

        \node [decision, below=3.4cm of list] (map) {
            Map?
        };

        \node [decision, right=2.0cm of map] (map-children) {
            Children?
        };

        \node [subroutine, right=1.5cm of map-children] (recur-map) {
            \nodepart{two}\shortstack{Recur\\on value}
        };

        \node [block, below of=recur-map] (map-child) {
            Map value to ID
        };

        \node [stop, above of=map-children] (return-node) {
            Return node
        };

        \node [decision, below of=map] (primitive) {
            Primitive?
        };

        \node [stop, below of=primitive] (error) {
            Error
        };
        
        
        \draw [line] (input.south) -- (list.north);
        \draw [line] (list.south) -- node [near start] {No} (map.north);
        \draw [line] (map.south) -- node [near start] {No} (primitive.north);
        \draw [line] (primitive.south) -- node {No} (error.north);
        
        \draw [line] (list.east) -- node {Yes} (list-children.west);
        \draw [line] (recur-list.west) -- (append-list.east);
        \draw [line] (append-list.south) -- (list-children.north);
        \draw [line] (list-children.east) -| node [near start] {Yes} (recur-list.south);
        \draw [line] (list-children.south) |- node [near start] {No} (return-node.east);
        
        \draw [line] (map.east) -- node [near start] {Yes} (map-children.west);
        \draw [line] (map-children.north) -- node [near start] {No} (return-node.south);
        \draw [line] (map-children.east) --  node [near start] {Yes} (recur-map.west);
        \draw [line] (recur-map.south) -- (map-child.north);
        \draw [line] (map-child.west) -| (map-children.south);
        \draw [line] (primitive.west) -| node [near start] {Yes} ($ (primitive.west) + (-1cm, 6cm) $) -- (return-node.west);
        
\end{tikzpicture}
}
\caption{A flowchart representing the translation of an abstract syntax tree node to an object node.} \label{fig:semantic}
\end{figure}

After parsing user syntax, Nanoverse constructs a partial representation of model semantics (Fig. \ref{fig:semantic}). This partial semantic model, known as the ``object node hierarchy'' or ``build hierarchy,'' encodes all requirements explicitly specified by the user. Nanoverse constructs this hierarchy through the use of a graph of symbol tables. 

At their most basic, symbol tables are mapping functions from a text symbol to some other value \cite{Grune2000}. A compiler for an imperative language will typically create a single symbol table at each level of contextual scope to associate identifiers with values \cite{Cooper2011}.  Nanoverse symbols, on the other hand, represent system components: i.e., loosely coupled subsystems that supply specific functionality for the containing system, and which themselves depend on further subsystems \cite{Bachmann2000}. As such, Nanoverse uses a symbol table for every component and component class. 

To encode its rule base, the Nanoverse compiler constructs two classes of symbol tables: \emph{resolving} symbol tables (RSTs) and \emph{instantiable} symbol tables (ISTs). RSTs narrow a particular identifier to a specific subclass of an expected class. ISTs resolve the names of specific subsystems required to instantiate an object of a specific class. Object translation proceeds by alternating between these two symbol table classes.

To begin translation, a root IST is supplied, along with the root node of the AST, to the translation visitor, or ``translator.'' The translator visits each child of the root node. If the identifier of the node does not exist in the IST, translation halts with an error. Assuming the symbol is recognized, the translator passes the AST node to the IST for resolution. The IST resolves the node against an RST, which provides another IST for translating the child. The child's IST and the child are then passed back to the translator, resulting in a depth-first traversal.

ISTs come in three varieties: list ISTs (LISTs), map ISTs (MISTs), and primitive instantiators. MISTs expect key-value pairs and correspond to single objects; LISTs expect multiple anonymous values and represent collections or predicate blocks. Primitive instantiators redesignate primitive AST nodes as object nodes. Translating a MIST involves resolving its identifier against the parent IST to retrieve an appropraite RST, then resolving its value against the RST to retrieve a child IST and calling back. Every element in a LIST has the same RST, so translating the LIST just involves resolving each LIST child node against the RST and calling back. Primitives are terminals; no callback is required.

After translation, the user's requirements have been translated into a hierarchy of constraints. The compiler must now determine whether and how a simulation can be instantiated from the user's specifications. The user's constraints may be expressly incompatible, or they may imply further requirements that are incompatible. In these cases, the model is overdetermined. On the other hand, the user may have omitted required fields (i.e., fields with no default values). If this happens, the model is underdetermined. Assuming neither an overdetermined nor underdetermined model, the compiler's next task is to interpolate sufficient constraints to fully determine the model's configuration.

\subsection{Interpolation and construction}

Nanoverse organizes a simulation into a hierarchy of components. ``Components'' in Nanoverse are equivalent to ``primitives'' in NetLogo or GAMA \cite{Wilensky2004,Grignard2013}, except that most components are not ``primitive'' in the sense of being discrete, atomic wholes. Rather, a Nanoverse component may have an arbitrary number of subcomponents, which may likewise have subcomponents of their own. Components are only loosely coupled to their subcomponents---often by a single method---facilitating interchange. Interchangeable components are at the heart of the configuration-based approach.

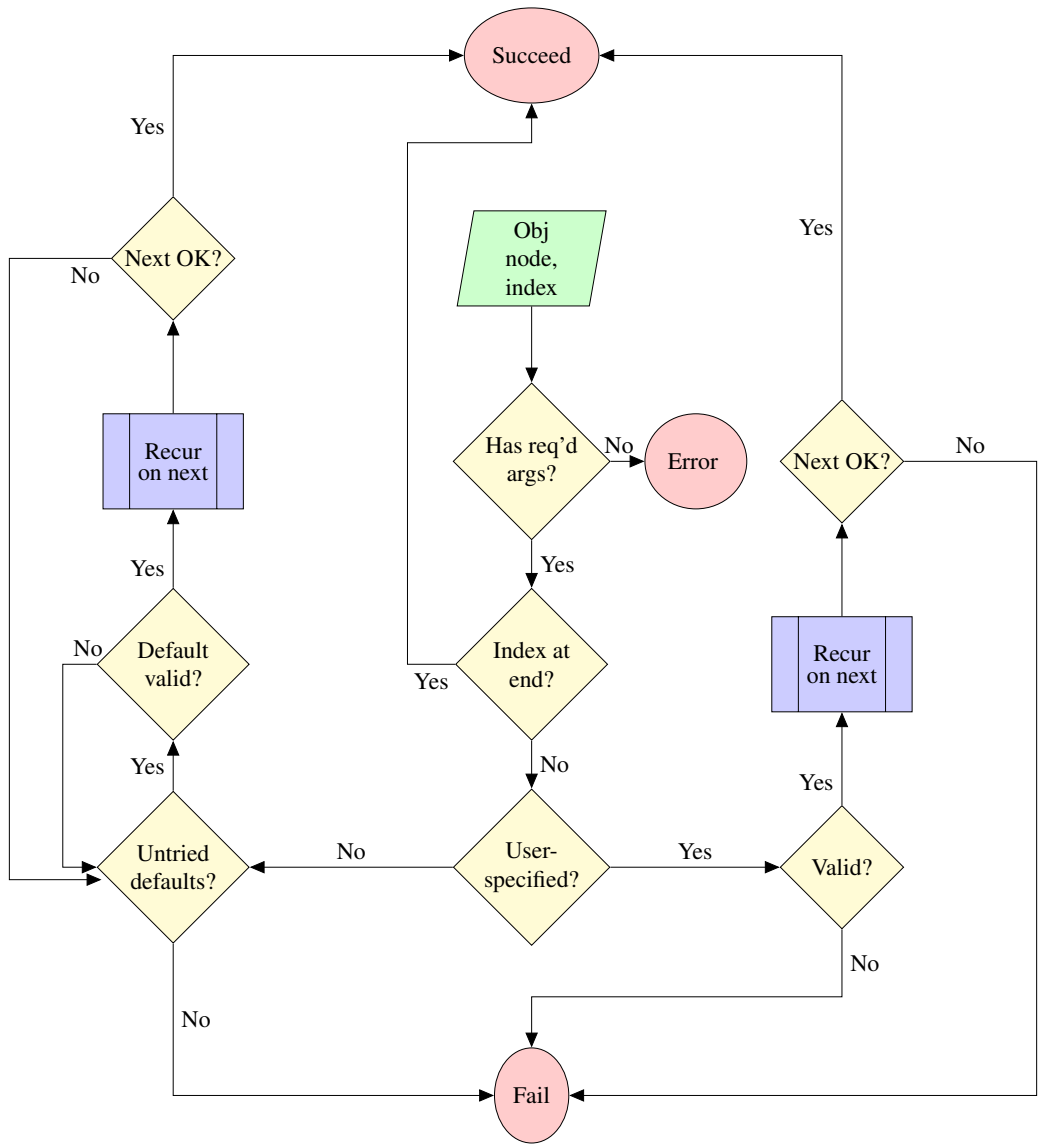
\begin{figure}[htb]
\centering
\scalebox{0.9}{
\begin{tikzpicture}[node distance = 3cm, auto]

		
        \node (succeed) [stop] at (0,0) {
			 Succeed
        };
        
        \node [data, text width=3.9em, below of=succeed] (input) {
            Obj node, index
        };
        
        \node [above left=1cm and 1cm of input] (hub1) {};
        
        \node [decision, below of=input] (required) {
            Has req'd args?
        };

        \node [stop, right=0.5cm of required, align=center] (error) {
            Error
        };

        \node [decision, below of=required] (end) {
            Index at end?
        };

        \node [decision, below of=end] (user) {
            User-specified?
        };

        \node [decision, left=3cm of user] (defaults) {
            Untried defaults?
        };
        
        \node [decision, above of=defaults] (default-valid) {
            Default valid?
        };
        
        \node [left=0.5cm of default-valid] (hub-default-valid) {};
        
        \node [subroutine, above of=default-valid] (default-recur) {
            \nodepart{two}\shortstack{Recur\\on next}
        };
        
        \node [decision, above of=default-recur] (default-next) {
            Next OK?
        };
        
        \node [left=1.5cm of default-next] (hub-default-next) {};
        \node [below left=8.6cm and 10pt of default-next] (hub-defaults) {};
        \node [decision, right=2.5cm of user] (user-valid) {
            Valid?
        };

        \node [subroutine, above of=user-valid] (user-recur) {
            \nodepart{two}\shortstack{Recur\\on next}
        };
        
        \node [decision, above of=user-recur] (user-next) {
            Next OK?
        };
        
        \node [right of=user-next] (hub-user-next) {};

        \node [stop, below=1.5cm of user] (fail) {
            Fail
        };

        \node [above=0.5cm of fail] (hub-fail-north) {};
        
		
		\draw [line] (input.south) -- (required.north);
		\draw [line] (required.east) -- node [near start] {No} (error.west);
		\draw [line] (required.south) -- node {Yes} (end.north);	
		\draw [line] (end.south) -- node {No} (user.north);
		\draw [line] (user.east) -- node {Yes} (user-valid.west);
		\draw [line] (user.west) -- node [above] {No} (defaults.east);
		\draw [line] (user-valid.north) -- node [near start] {Yes} (user-recur.south);
		\draw [line] (user-recur.north) -- (user-next.south);
		\draw [line] (user-next.north) |- node [near start] {Yes} (succeed.east);
		\draw [line] (end.west) -|  node [near start] {Yes} (hub1.south) -| (succeed.south);
		\draw [line] (user-next.east) -| node [near start] {No} (hub-user-next.west) |- (fail.east);
		\draw [line] (user-valid.south) |- node [near start] {No} (hub-fail-north.north) -- (fail.north);
		\draw [line] (defaults.north) -- node {Yes} (default-valid.south);
		\draw [line] (defaults.south) |- node [near start] {No} (fail.west);
		\draw [line] (default-valid.west) -- node [above, near start] {No} (hub-default-valid.east) |- (defaults.west);
		\draw [line] (default-valid.north) -- node [near start] {Yes} (default-recur.south);
		\draw [line] (default-recur.north) -- (default-next.south);
		\draw [line] (default-next.north) |- node [near start] {Yes} (succeed.west) ;
		\draw [line] (default-next.west) -- node [near start] {No} (hub-default-next.east) |- (hub-defaults.west);

\end{tikzpicture}
}
\caption{Flowchart representing the constraint satisfaction process used to interpolate unspecified user parameters.} \label{fig:backtrack}
\end{figure}

The configuration of Nanoverse components is accomplished through the hierarchical solution of local constraints. Each subcomponent has its own constraints. These constraints determine whether the component is compatible with the existing partial configuration, and which subcomponents can be supplied to it. Additionally, the subcomponents for a given component may depend on one another, and are thus supplied as additional constraints on the subcomponent. Associated with each subcomponent is one or more defaults, which are given in order of preference. Each default may imply its own set of constraints. If the user has specified a particular value for a subcomponent, the specified subcomponent (and its implied constraints) replaces the default list. Component configurations are then solved depth first until a total solution has been found, or it is determined that no solution exists (Fig. \ref{fig:backtrack}).

In a constraint satisfaction problem, solutions are often obtained through a backtracking scheme. A backtracking scheme consists of a recursive algorithm. Let $v_0,...v_i$ represent the values that must be specified, and let $D_i$ represent the domain of solutions for $v_i$. In addition, there exists a set of constraints on the solution set. The backtracker begins by seeking a value $v_0 = d_0 | d_0 \in D_0$ that satisfies the relation $C \cap v_0 = d_0$. If such a value is found, the algorithm recurs on $v_1, D_1$. For the $n$th recursion, the algorithm seeks a value of $v_n = d_n | d_n \in D_n$ that satisfies $C \cap \left(v_k = d_k \forall k \le n\right)$. If this relation is not satisfied, the algorithm returns failure \cite{Russell2009}.

In Nanoverse, the constraints represent the specific requirements of particular subcomponents. The constraint set $C$ is therefore not globally constant. However, for any given component, the only constraint is that all of its subcomponents are legal, given their dependencies. Thus, a subcomponent can verify constraint satisfaction by verifying that all of its subcomponents can find legal instance values. For simple subcomponents with only one possible default value, such a check is relatively straightfoward. More complex subcomponents must perform their own interpolation step. This component-dependent interpolation step is encapsulated in the ``Valid?'' decision node in Fig. \ref{fig:backtrack}.

Instantiation proceeds like interpolation. Each component has its own instantiation method, which builds any helper objects as necessary. For the most part, these helper objects are themselves components, albeit not user-specified ones. That is, they are only loosely coupled to the parent component, and they are automatically configured based on the properties of the parent component. Helper components include getters and setters from other runtime objects, which serve the same role as public method calls in traditional APIs.

\section{CHALLENGES AND FUTURE DIRECTIONS}

\subsection{Engineering considerations}

Component-based software design has been widely incorporated into many software platforms, most recently in the form of Java's ``Project Jigsaw'' \cite{Bachmann2000,Reinhold2014}. Difficulty of maintenance is a challenge that is common to all of these approaches \cite{Weyuker1998}. Component-based software essentially shifts some of the user's responsibilities onto the developer. Rather than define logic, the user builds pre-fabricated logical components into the desired configuration. The power and utility of a component-based platform is therefore limited by the breadth and quality of these pre-fabricated components. The developer must provide both the runtime logic and any steps required for the component to compile. 

In the case of Nanoverse, these steps include specifying acceptable sub-components and the order of preference for those components. This implies a larger codebase than analogous, imperative systems. Nanoverse is also fully declarative, unlike hybrid declarative-imperative languages like GAMA. As a fully declarative language, the only possible business logic is that which is defined in an existing component. Many agent-based models share some similar ideas, especially in Nanoverse's primary domain of spatially explicit ABMs. Special cases abound, however, and new questions lead constantly to new model designs. How can Nanoverse accommodate advanced use cases without complicating the simple ones? Perhaps the most straightforward solution is to incorporate an imperative sub-language, which would be backed by a simple interpreter. This would have the added benefit of allowing model changes on the fly, and even self-modifying code, which can be used for evolutionary simulations.

Testing is another major challenge. It is often desirable to test a component both in isolation and in its intended context. However, the number of possible contexts for any given component is limitless: as user models (and the component library) grow, the same component can be nested deep in a hierarchy of other parts. When the assumptions of components are in contradiction, unexpected behavior can result. These risks can be managed through the judicious use of consistency checks, strong interface contracts, and exhaustive unit testing \cite{Crnkovic2002}.

\subsection{Default generality}

In the Nanoverse prototype, the only planned constraints have to do with logical compatibility. For example, a spatially explicit model taking place in a hexagonal arena cannot employ a periodic boundary condition, because two of the six sides would remain unmatched. Likewise, a rectangular lattice cannot employ a hexagonal arena. This lowers the skill threshold required for use, but it does not handle another important class of constraint: preventing the selection of many parameters that, while technically not in conflict, may produce unexpected behavior.

There are many situations in which the user would expect different defaults based on his or her selections, even if one default could technically satisfy all cases. This is particularly true for spatially structured systems. Consider, for example, the resolution of collisions. What should happen if an agent that is scheduled to move has no vacant space into which it can go? The user must specify how to choose a destination, and, if collisions are possible, how to resolve them. If the user specifies that destinations must include occupied locations, there must be a rule for resolving a collision. Conversely, if the user specifies a rule for resolving collisions, it is likely that the user wishes to allow them. That said, ``ignore occupied spaces'' remains a valid default behavior for the rule. Likewise, permitting occupied spaces is compatible with a resolution strategy of ``throw an error on collisions,'' though this is unlikely to be desired.

Within a specific application area, there may also be practical constraints on default behavior that relate to the meaning of the model. For example, the preferred default resolution of collisions in a forest fire model (intensify the fire) is different from that of a microbial model (push the existing occupant away). One solution is the ability to specify custom constraints and default sets, and to inherit these elements as domain-specific libraries. For the previous example, one might create a default component hierarchy for ``bacteria'' agents, and another for ``flame'' agents.

\section{CONCLUSION}

The Nanoverse compiler has the potential to simplify the process of building agent-based models. This greater ease can benefit both novice and experienced users, as fewer parameters need be considered or specified. With the introduction of component libraries, Nanoverse can also function as a medium for the transmission of expert knowledge concerning model design. As with many agent-based modeling platforms, the same approach can be used to simplify the design of interactive systems, such as games.

The strict hierarchical structure of the Nanoverse language provides several benefits. The Nanoverse compiler already exploits the most important of these: the availability of algorithms to interpolate missing nodes. Hierarchies are also scale-free, which makes them easy to visualize, e.g. using a zooming user interface \cite{Bederson1998}. The strict separation of concerns required for hierarchical design also simplifies compiler design, which facilitates optimization of program flow. Finally, a component-based approach facilitates the automatic generation of documentation via traversal of the symbol table hierarchy.

The nanoverse compiler is under active development. Our first goal is to port all runtime functionality from the interpreted Nanoverse prototype \cite{Borenstein2015}, including modular topology and continuum-valued fields, to the compiler-based edition. Following that, we plan to provide an automatic documentation system and publish it to the Nanoverse website. Beyond that, we will focus on addressing the limitations of the language by introducing user-defined variables, user-defined constraints and defaults, object orientation, and code importation. 

\section*{ACKNOWLEDGMENTS}

The author gratefully acknowledges Ned S. Wingreen for his support and guidance throughout the development of the Nanoverse framework, as well as Anne Maslan for her helpful feedback and testing.

\bibliographystyle{unsrturl}
\bibliography{arXiv2}

\end{document}